\documentclass[prd,eqsecnum,twocolumn,amsfonts,amssymb]{revtex4}

\usepackage{graphicx}

\usepackage{bm}
\usepackage{framed}

\newcommand{\beq}{\begin{equation}}
\newcommand{\eeq}{\end{equation}}
\newcommand{\beqs}{\begin{eqnarray}}
\newcommand{\eeqs}{\end{eqnarray}}
\newcommand{\lsim}{\mathrel{\raisebox{-
.6ex}{$\stackrel{\textstyle<}{\sim}$}}}

\begin{document}

\title{Anomalous Dimensions at an Infrared Fixed Point
  in an SU($N_c$) Gauge Theory with Fermions in the
  Fundamental and Antisymmetric Tensor
  Representations}

\author{Thomas A. Ryttov$^a$ and Robert Shrock$^b$}

\affiliation{(a) \ CP$^3$-Origins, University of Southern Denmark, \\
Campusvej 55, Odense, Denmark}

\affiliation{(b) \ C. N. Yang Institute for Theoretical Physics and 
Department of Physics and Astronomy, \\
Stony Brook University, Stony Brook, NY 11794, USA }

\begin{abstract}

  We present scheme-independent calculations of the anomalous
  dimensions $\gamma_{\bar\psi\psi,IR}$ and $\gamma_{\bar\chi\chi,IR}$
  of fermion bilinear operators $\bar\psi\psi$ and $\bar\chi\chi$ at
  an infrared fixed point in an asymptotically free SU($N_c$) gauge
  theory with massless Dirac fermion content consisting of $N_F$
  fermions $\psi^a_i$ in the fundamental representation and $N_{A_2}$
  fermions $\chi^{ab}_j$ in the antisymmetric rank-2 tensor
  representation, where $i,j$ are flavor indices. For the case
  $N_c=4$, $N_F=4$, and $N_{A_2}=4$, we compare our results with
  values of these anomalous dimensions measured in a recent lattice
  simulation and find agreement.
    
\end{abstract}

\maketitle

% =======================================================================

% section 1
\section{Introduction}
\label{intro_section}

An asymptotically free gauge theory with sufficiently many massless
fermions has an infrared zero in its beta function, which is an
infrared fixed point (IRFP) of the renormalization group
\cite{b1,b2,bz}.  At this IRFP the theory is scale-invariant and is
inferred to be conformally invariant \cite{scalecon},
whence the commonly used term ``conformal window'' (CW).  Because of
the asymptotic freedom, one can use perturbation theory reliably in
the deep ultraviolet (UV) where the gauge coupling approaches zero,
and then follow the renormalization-group flow toward the infrared. These
statements apply to both vectorial and chiral gauge theories; here we
restrict our consideration to vectorial gauge theories.  As the
fermion content is reduced, the gauge coupling at the IRFP increases
in strength and eventually exceeds a value such that there is
generically spontaneous chiral symmetry breaking and dynamical fermion
mass generation.  This defines the lower end of the conformal
window. Theories that lie slightly below this lower end exhibit
quasi-conformal behavior over a large interval of Euclidean
energy/momentum scales, over which the gauge coupling runs slowly due
to a small beta function. Such theories just below the lower end of
the conformal window can be relevant to approaches to composite-Higgs
scenarios and associated physics beyond the Standard Model (BSM).
Considerable progress has been made in studies of quasi-conformal
vectorial gauge theories with several flavors of fermions
transforming according to a single representation of the gauge group,
e.g., SU(3) with $N_F=8$ Dirac fermions in the fundamental
representation \cite{lsd_2016}-\cite{lsd_2019}.

For an operator ${\cal O}$, the full scaling dimension is denoted as
$D_{\cal O}$ and its free-field value as $D_{{\cal O},free}$.  The
anomalous dimension of this operator, denoted $\gamma_{\cal O}$, is
defined via the relation $D_{\cal O} = D_{{\cal O},free} -
\gamma_{\cal O}$.  The anomalous dimensions of gauge-invariant
operators at an IRFP are of basic physical interest. While the
simplest gauge theories have used fermions transforming according to a
single representation of the gauge group, a natural generalization is
to study theories with multiple fermions transforming according to
different representations of the gauge group. In previous work
\cite{dexm}, we presented scheme-independent perturbative calculations
of anomalous dimensions of fermion bilinears at an IRFP in the
conformal window in a theory of this type (in $d=4$ spacetime
dimensions at zero temperature), with a general non-Abelian gauge
group $G$ and massless fermion content consisting of $N_f$ fermions
$f$ in a representation $R$ and $N_{f'}$ fermions $f'$ in a
representation $R'$ of $G$ \cite{fm}.  Our calculational method
applies at an exact IRFP in the conformal window (sometimes called the
non-Abelian Coulomb phase).  In \cite{dexml} with S. Girmohanta we
studied theories of this type with $G={\rm SU}(N_c)$, $R$ equal to the
fundamental representation, and $R'$ equal to the adjoint or rank-2
tensor representation and investigated a type of 't Hooft-Veneziano
limit, $N_f \to \infty$, $N_c \to \infty$ with $N_f/N_c$ and $N_{f'}$
fixed.

We denote a gauge theory with gauge group $G={\rm SU}(N_c)$ and
(massless) fermion content consisting of $N_F$ Dirac fermions in the
fundamental representation, denoted $F$, and $N_{A_2}$ Dirac fermions
in the antisymmetric rank-2 tensor representation, denoted $A_2$, as
$(N_c,N_F,N_{A_2})$ for short.  Recently, Ref. \cite{fx444} has
reported interesting results from lattice simulations of the
$(N_c,N_F,N_{A_2})=(4,4,4)$ theory. Since the $A_2$ representation in
SU(4) is self-conjugate, the $N_{A_2}$ Dirac fermions are equivalent
to $2N_{A_2}$ Majorana fermions.  Ref. \cite{fx444} finds evidence for
an IRFP inferred to lie in the conformal window, near its lower end,
and presents measurements of the anomalous dimensions of the $F$ and
$A_2$ fermion bilinears $\gamma^{(4)}_m$ and $\gamma^{(6)}_m$ (where
the superscripts refer to the dimensionalities of these
representations of SU(4)), and of gauge-singlet composite-fermion
operators.  Given the conclusion in \cite{fx444} that this theory is
in the conformal window, a relevant question is whether our general
higher-order perturbative calculations of anomalous dimensions of
fermion bilinears in \cite{dexm}, when specialized to this theory,
yield results in agreement with the values measured in
\cite{fx444}. To our knowledge, this question has not been previously
investigated in the literature. 

In the present work we address and answer this question for the
anomalous dimensions $\gamma^{(4)}_m$ and $\gamma^{(6)}_m$ by
extracting the requisite special case of our general calculations of
anomalous dimensions of fermion bilinears in \cite{dexm} for the
$(N_c,N_F,N_{A_2})=(4,4,4)$ theory. To state our conclusions in
advance, to within the uncertainties in our finite-order perturbative
calculation, we find agreement with the lattice results in
\cite{fx444}.  The authors of Ref. \cite{fx444} also observed that
their conclusion that the (4,4,4) theory is in the conformal window
disagreed with a calculation in \cite{khl} (denoted KHL) of the lower
boundary of this conformal window in the $(4,N_F,N_{A_2})$ theory
based on a critical condition on ${\rm
  max}(\gamma_m^{(4)},\ \gamma_m^{(6)})$, denoted $\gamma$CC.  We
investigate this further here. 

As noted above, our calculations of anomalous dimensions of fermion
bilinears in \cite{dexm} were for a general non-Abelian gauge group
$G$ and fermion representations $R$ and $R'$. Before presenting our
calculations for SU(4) theory, we will first specialize the results
from \cite{dexm} to the case of the gauge group $G={\rm SU}(N_c)$ with
massless fermion content consisting of $N_F$ Dirac fermions in the
fundamental representation and $N_{A_2}$ Dirac fermions in the
antisymmetric rank-2 tensor representation of SU($N_c$), i.e.,
theories of the type $(N_c,N_F,N_{A_2})$ in our shorthand notation.
We then further specialize to the case $N_c=4$, 
and then finally to the $(4,4,4)$ theory.

We denote the massless Dirac fermions in the $F$ and $A_2$
representations as $\psi^a_i$ and $\chi^{ab}_j=-\chi^{ba}_j$, where
$a,b$ are SU($N_c$) gauge indices, and $i,j$ are flavor indices
with $i=1,...,N_F$ and
$j=1,...,N_{A_2}$.  We shall use our general
results in \cite{dexm} to calculate scheme-independent series
expansions for the anomalous dimensions at the IRFP, denoted
$\gamma_{\bar\psi\psi,IR}$ and $\gamma_{\bar\chi\chi,IR}$, of the
respective (gauge-invariant) fermion bilinears
\beq
\bar\psi\psi = \sum_{i=1}^{N_f} \bar\psi_{a,i} \psi^a_i
\label{psibarpsi}
\eeq
and 
\beq
\bar\chi\chi = \sum_{j=1}^{N_{A_2}} \bar\chi_{ab,j} \chi^{ab}_j \ ,
\label{chibarchi}
\eeq
where the sums over color indices are understood and run from 1 to
$N_c$.  Although we take the fermions to be massless, the
operators (\ref{psibarpsi}) and (\ref{chibarchi}) would be mass
operators (with all fermion flavors taken to have equal mass) if these
fermions were massive, and for this reason another common notation
for the anomalous dimensions is $\gamma^{(F)}_m \equiv
\gamma_{\bar\psi\psi,IR}$ and $\gamma^{(A_2)}_m \equiv
\gamma_{\bar\chi\chi,IR}$. In the special case $N_c=4$, these are also
written with reference to the respective dimensionalities 4 and 6 of
the $F$ and $A_2$ representations as $\gamma^{(4)}_m \equiv
\gamma_{\bar\psi\psi,IR}$ and $\gamma^{(6)}_m \equiv
\gamma_{\bar\chi\chi,IR}$.  The color and flavor indices will often be
suppressed in the notation.

It should be mentioned that studies were also performed of the (4,2,2)
theory, due to its possible role as a model for a composite Higgs
boson and a partially composite top quark
\cite{degrand_etal}-\cite{deldebbio}.  However, as we noted in
\cite{dexm}, the (4,2,2) theory is in the chirally broken phase, where
there is no exact IRFP and hence where our calculations are not
directly applicable. In general, BSM theories with fermions in
higher-dimensional representations have long been of interest; in
addition to Refs. \cite{fx444} and
\cite{degrand_etal}-\cite{deldebbio}, some of the many works include
\cite{higherrep,sextet,ferretti}.

This paper is organized as follows.  In Section
\ref{calculations_section} we briefly discuss some relevant background
and our general calculational methods. Section
\ref{general_results_section} contains our results for the anomalous
dimensions for SU($N_c$) with general $N_F$ and $N_{A_2}$ in the
conformal window, while Section \ref{su4_section} presents the
corresponding formulas for $N_c=4$ and for the specific
$(N_c,N_F,N_{A_2})=(4,4,4)$ theory. 
Our conclusions are given in Section \ref{conclusions_section}.

% ===================================================================

\section{Calculational Methods}
\label{calculations_section}

In this section we briefly review our calculational methods and
relevant notation.  In the context of the general SU($N_c$) gauge
group, we first mention two degenerate cases. If $N_c=2$, then the
$A_2$ representation is a singlet and hence decouples from the
dynamics, so the theory reduces to one with fermions in just the
fundamental representation.  Hence, in all expressions involving the
number $N_{A_2}$, this number always occurs multipled by the factor
$(N_c-2)$.  If $N_c=3$, then $A_2 = \bar F$, i.e., the $A_2$
representation is the conjugate fundamental representation of
SU(3). Taking into account the fact that a Dirac fermion $f$ has a
decomposition into chiral components $f= f_L+f_R$ and the property
that a left-handed chiral component of a fermion can be equivalently
written as the charge conjugate of a right-handed antifermion, it
follows that if $N_c=3$, then the theory reduces to one with
$N_F+N_{A_2}$ fermions in the fundamental representation. If $N_c \ge
4$, then the $F$ and $A_2$ representations are distinct.

% ======================================================

\subsection{Relevant Range of $N_F$ and $N_{A_2}$}

We denote the running gauge coupling as $g=g(\mu)$, where $\mu$ is the
Euclidean energy/momentum scale at which this coupling is measured.
We define $\alpha(\mu) = g(\mu)^2/(4\pi)$.  As noted before, since we require
the theory to be asymptotically free, its properties can be computed
perturbatively in the UV limit at large $\mu$, where $\alpha(\mu) \to 0$.
The dependence of $\alpha(\mu)$ on $\mu$ is
described by the renormalization-group (RG) beta function,
\beq
\beta = \frac{d\alpha(\mu)}{d\ln\mu} \ .
\label{beta}
\eeq
The argument $\mu$ will generally be suppressed
in the notation.  The series expansion of $\beta$ in powers of
$\alpha$ is
\beq
\beta = -2\alpha \sum_{\ell=1}^\infty b_\ell \, a^\ell \ ,
\label{beta_series}
\eeq
where
\beq
a \equiv \frac{g^2}{16\pi^2} = \frac{\alpha}{4\pi} \ , 
\label{a}
\eeq
and $b_\ell$ is the
$\ell$-loop coefficient.  For a theory with a gauge group $G$ and Dirac
fermions $f$ and $f'$ in respective representations $R$ and $R'$ of $G$,
the one-loop coefficient in the beta function is \cite{b1}
\beq
b_1 = \frac{1}{3}\Big [ 11C_A - 4(N_fT_f + N_{f'}T_{f'}) \Big ] \ , 
\label{b1}
\eeq
and the two-loop coefficient is \cite{b2}
\beq
b_2 = \frac{1}{3}\bigg [ 34C_A^2 - 4N_fT_f(5C_A+3C_f)
  -4N_{f'}T_{f'}(5C_A+3C_{f'}) \bigg ] \ , 
\label{b2}
\eeq
where $C_A$, $C_f$, and $T_f$ are group invariants
(see \cite{rsv} and the Appendix). 
With an overall minus sign extracted, as in Eq. (\ref{beta}), the
condition of asymptotic freedom is that $b_1 > 0$. Setting $R=F$ and
$R'=A_2$ and substituting the values of the group invariants for
$G={\rm SU}(N_c)$, the condition $b_1 > 0$ reads
\beq
N_F + (N_c-2)N_{A_2} < \frac{11N_c}{2} \ .
\label{b1positive}
\eeq
The resultant upper ($u$) limits on $N_F$ and $N_{A_2}$ imposed by
the requirement of asymptotic freedom are thus 
\beq
N_{F,u} = \frac{11}{2}N_c - (N_c-2)N_{A_2}
\label{Nfundup}
\eeq
and
\beq
N_{A_2,u} = \frac{11N_c - 2N_F}{2(N_c-2)} \ .
\label{Nasymup}
\eeq
The maximal order to which the beta function is independent of the
scheme used for regularization and renormalization is the two-loop
order. With $b_1 > 0$, the condition that this two-loop beta function
should have an IR zero is that $b_2 < 0$.  For the SU($N_c$) theory,
this is the condition 
\beq
N_F(13N_c^2-3)+2N_{A_2}(N_c-2)(8N_c^2-3N_c-6) > 34N_c^3 \ . 
\label{b2negative}
\eeq
The region of the first quadrant in the $(N_F,N_{A_2})$ plane where
the inequalities (\ref{b1positive}) and (\ref{b2negative}) are both
satisfied will be denoted $I_{IRZ}$, where the subscript refers to the
existence of an IR zero (IRZ) in the beta function.  We label the
upper and lower boundaries of the $I_{IRZ}$ region as ${\cal
  B}_{IRZ,u}$ and ${\cal B}_{IRZ,\ell}$, respectively. In plotting
these boundaries, one formally generalizes $N_F$ and $N_{A_2}$ from
positive integers (or half-integers for $N_{A_2}$ if $N_c=4$) to
positive real numbers, with the understanding that the physical cases
are integral (or half-integral for $N_{A_2}$ if $N_c=4$).
Analogously, we denote the upper and lower boundaries of the conformal
window as ${\cal B}_{CW,u}$ and ${\cal B}_{CW,\ell}$,
respectively. The upper boundary ${\cal B}_{CW,u} = {\cal B}_{IRZ,u}$
is the solution locus to the condition $b_1=0$. The lower boundary
${\cal B}_{CW,\ell}$ is not exactly known even for the case of
theories with fermions transforming in only one representation;
indeed, much work using lattice simulations has been, and continues to
be, devoted to determining the approximate location of this lower
conformal-window boundary \cite{lgtreviews,simons}.  We discuss this lower
boundary ${\cal B}_{CW,\ell}$ further below for the $N_c=4$ theory.

Following our labelling convention in \cite{dexm}, we take the
horizontal and vertical axes of the first quadrant of the
$(N_F,N_{A_2})$ plane to be the $N_F$ and $N_{A_2}$ axes,
respectively.  The boundaries of the $I_{IRZ}$ region given by the equations
$b_1=0$ and $b_2=0$ are both line segments in this first quadrant of
the $(N_F,N_{A_2})$ plane.  In general, the slope of the line $b_1=0$
is
\beq
\frac{dN_{A_2}}{dN_F} = -\frac{1}{N_c-2} \ , 
\label{b1zero_line_slope}
\eeq
and the slope of the line $b_2=0$ is
\beq
\frac{dN_{A_2}}{dN_F} =
-\frac{(13N_c^2-3)}{2(N_c-2)(8N_c^2-3N_c-6)} \ .
\label{b2zero_line_slope}
\eeq
%

% ===================================================

\subsection{Higher-Order Terms in Beta Function}

 For a theory with a general gauge group $G$ and $N_f$ fermions in a
 single representation, $R$, the coefficients $b_1$ and $b_2$ were
 calculated in \cite{b1} and \cite{b2}, while $b_3$, $b_4$, and $b_5$
 were calculated in the commonly used $\overline{\rm MS}$ scheme
 \cite{msbar} in
 \cite{b3a,b3b}, \cite{b4}, and \cite{b5}, respectively (see also
 \cite{b5su3}). For the analysis of a theory with fermions in multiple
 different representations, one needs generalizations of these
 results.  These are straightforward to derive in the case of $b_1$
 and $b_2$, but new calculations are required for higher-loop
 coefficients.  These were performed in \cite{zoller}
 (again in the $\overline{\rm MS}$ scheme) up to four-loop order, and
 we used the results of Ref. \cite{zoller} in \cite{dexm}.

% ===================================================

\subsection{Anomalous Dimensions}

The conventional expansion of the anomalous dimension
$\gamma_{\bar f f}$ of the fermion bilinear $\bar f f$ in a gauge theory,
in terms of the squared gauge coupling, is 
\beq
\gamma_{\bar f f} = \sum_{\ell=1}^\infty c^{(f)}_\ell \, a^\ell \ ,
\label{gamma_series}
\eeq
where $c^{(f)}_\ell$ is the $\ell$-loop coefficient, and
correspondingly for $\gamma_{\bar {f'},f'}$ in a theory with both $f$
and $f'$ fermions.  These expansions apply, in particular, at an IRFP.
They may also be useful in the analysis of a quasi-conformal field
theory with parameters such that it lies slightly below the lower end
of the conformal window and hence exhibits UV to IR evolution over an
extended interval of $\mu$ governed by an approximate IRFP.  The
one-loop coefficient $c^{(f)}_1$ is scheme-independent, while the
$c^{(f)}_\ell$ with $\ell \ge 2$ are scheme-dependent, and similarly
with the $c^{(f')}_\ell$.  For a general gauge group $G$ and $N_f$
fermions in a single representation $R$ of $G$, the $c^{(f)}_\ell$
have been calculated up to loop order $\ell=4$ in \cite{c4a,c4b} and
$\ell=5$ in \cite{c5}. For the case of multiple fermion
representations, the coefficients $c^{(f)}_\ell$  
have been calculated up to four-loop order in
\cite{chetzol} in the $\overline{\rm MS}$ scheme.

Physical quantities such as anomalous dimensions at an IRFP clearly
must be scheme-independent. In conventional computations of these
quantities, one first writes them as series expansions in powers of
the coupling as in (\ref{gamma_series}), and then evaluates these
series expansions with $\alpha$ set equal to $\alpha_{IR}$, calculated
to a given loop order.  These calculations have been performed for
anomalous dimensions of gauge-invariant fermion bilinears in a theory
with a single fermion representation up to the four-loop level
\cite{bvh}-\cite{bc} and to the five-loop level in \cite{flir}.
However, as is well known, these conventional (finite-order) series
expansions are scheme-dependent beyond the leading terms.  Studies of
scheme dependence in the context of an IRFP have been carried out in
\cite{sch}-\cite{schemegauge}.  The fact that the conventional series
expansions for physical properties are scheme-dependent does not, by
itself, reduce the usefulness of these expansions. For example, this
scheme dependence is also true of higher-order calculations in quantum
chromodynamics (QCD), which were used to analyze data from hadron
colliders such as the Tevatron at Fermilab and the Large Hadron
Collider at CERN. Considerable effort has been, and continues to be,
expended to construct and apply schemes that minimize higher-order
contributions in these QCD calculations \cite{brodsky}.  Indeed, in
QCD, because the RG fixed point is an ultraviolet fixed point at the
origin in coupling constant space, it is, in principle, possible to
transform to a scheme where the beta function has no terms higher than
two-loop order (the 't Hooft scheme) \cite{hooft77}.  However, as was
shown in \cite{sch,sch23}, it is considerably more difficult to try to carry
out such a scheme transformation to remove terms at loop order 3 and
higher for a fixed point away from the origin.

Thus, in the analysis of the properties of a theory at a fixed point
away from the origin, as in the case of the IRFP of interest here, it
is useful to employ a series expansion method for calculating physical
quantities, such as anomalous dimensions, with the property that the
results to each order are scheme-independent.  A simple fact makes
this possible: at the upper end of the conformal window, as $b_1 \to
0$, this implies that $\alpha_{IR} \to 0$. Hence, one can reexpress a
series expansion at an IRFP in the conformal window as an expansion in
the manifestly scheme-independent variable $b_1$. For a theory with
fermions $f$ in a single representation, it is natural to use the
scheme-independent Banks-Zaks variable \cite{bz,gkgg} $\Delta_f =
N_{f,u}-N_f$.  Such calculations were carried out in
\cite{gtr}-\cite{baryon}.

In \cite{dexm} we generalized this analysis to theories with fermions $f$
and $f'$ in different representations $R$ and $R'$ of a general gauge group,
$G$.  The corresponding expansion variables for the scheme-independent
series expansions of physical quantities at an IRFP are
\beqs
\Delta_f &=& N_{f,u} - N_f \cr\cr
&=& \frac{11 C_A - 4N_{f'}T_{f'} - 4N_fT_f}{4T_f} \cr\cr
&=& = \frac{3b_1}{4T_f} \ , 
\label{Deltafgen} 
\eeqs
and similarly for $\Delta_{f'}$ with $f \leftrightarrow f'$.
Note that these expansion variables satisfy the relation 
\beq
\Delta_{f'} = \frac{T_f}{T_{f'}} \, \Delta_f \ .
\label{Delta_rel}
\eeq
The scheme-independent expansion for $\gamma_{\bar f f,IR}$ is 
\beq
\gamma_{\bar f f, IR} = \sum_{j=1}^\infty \kappa^{(f)}_j \, \Delta_f^j \ , 
\label{gamma_IR_Deltaseries}
\eeq
and similarly for $\gamma_{\bar{f'}f',IR}$ with $f \to f'$.  The
calculation of the coefficient $\kappa^{(f)}_j$ in
Eq. (\ref{gamma_IR_Deltaseries}) requires, as inputs, the values of
the $b_\ell$ in Eq. (\ref{beta_series}) for $1 \le \ell \le j+1$ and
the $c_\ell$ for $1 \le \ell \le j$.  We refer the reader to our
previous papers for further details of the calculations. 

Using the calculation of the beta
function for multiple fermion representation to four-loop order in
\cite{zoller}, together with the calculation in \cite{chetzol} of the
anomalous dimension coefficients in (\ref{gamma_series}) up to $\ell=3$ loop
order, we can calculate $\gamma_{\bar f f IR}$ to order
$O(\Delta_f^3)$ and $\gamma_{\bar {f'} f',IR}$ to order
$O(\Delta_{f'}^3)$  Parenthetically, note that we
cannot make use of the four-loop calculation of the $c^{(f)}_\ell$ 
in \cite{chetzol} to compute
$\gamma_{\bar f f,IR}$ to order $O(\Delta_f^4)$ and
$\gamma_{\bar {f'} f',IR}$ to $O(\Delta_{f'}^4)$, because this would
require, as an input, the five-loop coefficient $b_5$ in the beta
function for this case of multiple fermion representations, and, to
our knowledge, this has not been calculated.

In our specific application here, where the $f$ and $f'$ fermions transform
according to the representations $R=F$ and $R'=A_2$ of SU($N_c$), we will 
write these scheme-independent series expansions as 
\beq
\gamma_{\bar\psi\psi,IR} = \sum_{j=1}^\infty \kappa^{(F)}_j \Delta_F^j
\label{gamma_fund_Deltaseries_fasym}
\eeq
and
\beq
\gamma_{\bar\chi\chi,IR} = \sum_{j=1}^\infty \kappa^{(A_2)}_j
\Delta_{A_2}^j \ ,
\label{gamma_asym_Deltaseries_fasym}
\eeq
where
\beq
\Delta_F = \frac{11}{2}N_c - N_F - (N_c-2)N_{A_2}
\label{Delta_F}
\eeq
and
\beq
\Delta_{A_2} = \frac{11N_c - 2N_F - 2(N_c-2)N_{A_2}}{2(N_c-2)} \ . 
\label{Delta_{A2}}
\eeq
For this $(N_c,N_F,N_{A_2})$ theory, Eq. (\ref{Delta_rel}) reads 
\beq
\Delta_{A_2} = \frac{T_F}{T_{A_2}} \, \Delta_F = \frac{\Delta_F}{N_c-2} \ .
\label{Delta_rel_fasym}
\eeq
The truncation of the series (\ref{gamma_fund_Deltaseries_fasym}) to order
$O(\Delta_F^p)$ is denoted as $\gamma_{\bar\psi\psi,IR,\Delta_F^p}$ and
similarly, the truncation of the series (\ref{gamma_asym_Deltaseries_fasym})
to order $O(\Delta_{A_2}^p)$ is denoted
$\gamma_{\bar\chi\chi,IR,\Delta_{A_2}^p}$.  In accord with the remarks on
the $N_c=3$ special of the theory at the beginning of this section, we
note the identities
\beqs
N_c = 3 \ & \Rightarrow& \kappa^{(F)}_j = \kappa^{(A_2)}_j \ , \cr\cr
&& \Delta_F = \Delta_{A_2} \ , \cr\cr
&& \gamma_{\bar\psi\psi,IR,\Delta_F^p} =
\gamma_{\bar\chi\chi,IR,\Delta_{A_2}^p} \ .
\label{su3_identities}
\eeqs

In general, series expansions in powers of interaction couplings in
quantum field theory are asymptotic expansions rather than Taylor
series. As we discussed in \cite{pgb}, the scheme-independent
expansion (\ref{gamma_IR_Deltaseries}) is also generically an
asymptotic expansion rather than a Taylor series expansion with finite
radius of convergence. This is a consequence of the property that in
order for a series expansion of a function $f(z)$ in powers of $z$ to
be a Taylor series with finite radius of convergence, it is necessary
(and sufficient) that $f(z)$ must be analytic at the origin of the
complex $z$ plane. With $z = \Delta_f$, this means that the properties
of the theory should remain qualitatively similar for small positive
and negative real $\Delta_f$.  However, as $\Delta_f$ passes from real
positive values through zero to negative real values, i.e., as $N_f$
increases through the value $N_{f,u}$, the theory changes
qualitatively from being asymptotically free to being
IR-free. Nevertheless, just as with perturbative calculations in
quantum electrodynamics, one may still use the scheme-independent
expansions (\ref{gamma_fund_Deltaseries_fasym}) and
(\ref{gamma_asym_Deltaseries_fasym}) to get approximate information
about these anomalous dimensions. In our previous works, e.g.,
\cite{gtr,gsi,dex,dexl,dexo,pgb}, we have carried out the requisite
assessment of higher-order contributions, up to order $O(\Delta_f^4)$
for $\gamma_{\bar f f,IR}$ and $O(\Delta_f^5)$ for $\beta'_{IR}$ in
theories with fermions in a single representation. These showed that
the scheme-independent series expansions are reasonably convergent
throughout the conformal window, although, of course, the higher-order
terms make relatively larger contributions as one approaches the lower
end of this window.  The curves that we will show below for
$\gamma_{\bar\psi\psi,IR,\Delta_F^p}$ and
$\gamma_{\bar\chi\chi,IR,\Delta_{A_2}^p}$ for $1 \le p \le 3$ provide
an analogous quantitative measure of the effective convergence of
these expansions.

Interestingly, in \cite{dexss} we studied the ${\cal N}=1$
supersymmetric SU($N_c$) theory with matter content consisting of
$N_F$ copies of chiral superfields and their conjugates, for which the
anomalous dimension of the gauge-invariant chiral superfield bilinear
is exactly known \cite{nsvz,seiberg}, and we showed (a) that the
$\kappa_j$ coefficients precisely reproduce the series
expansion coefficients of the exact results to all orders, and (b) the
scheme-independent expansion of this anomalous dimension is convergent
throughout the full nonabelian Coulomb phase, which corresponds to the
conformal window in that theory.

% ==============================================================

\subsection{Condition on Anomalous Dimensions for Conformal Window}
\label{gamma_crit}

On the basis of analyses of the Schwinger-Dyson equation for the
propagator of a fermion $f$, operator product expansions, and other
arguments \cite{alm,cohen_georgi,kaplan_etal,zwicky}, it has been
suggested that an upper bound
\beq
\gamma_{\bar f f, IR} \le 1
\label{gamma_le_1}
\eeq
applies for an IRFP in the conformal window. In view of the
uncertainties pertaining to strong coupling and nonperturbative
effects, this bound is also sometimes stated as $\gamma_{\bar f f, IR}
\lsim 1$; here we will take this as implicit in our discussions.
Since $\gamma_{\bar f f,IR}$ increases as one moves down through the
conformal window from the upper end where $b_1=0$, it follows that
when the inequality (\ref{gamma_le_1}) is saturated, i.e, when the
critical (denoted $\gamma$CC)
\beq
\gamma_{\bar f f,IR}=1
\label{gamma_eq_1}
\eeq
holds, this defines the lower end of the conformal window, ${\cal
  B}_{CW,\ell}$. As we discussed in \cite{dexss}, this is true in the
case of an ${\cal N}=1$ supersymmetric theory with gauge group
SU($N_c$) and a set of chiral superfields in the $F$ and $\bar F$
representations, where the anomalous dimension of the gauge-invariant
chiral superfield bilinear is exactly known \cite{nsvz,seiberg}.  The
occurrence of the quadratic equation
\beq
\gamma_{\bar f f,IR}(2-\gamma_{\bar f f,IR})=1
\label{gamma_quadratic}
\eeq
as a critical condition for fermion condensation and its connection
with the condition (\ref{gamma_eq_1}) was noted in \cite{alm}.  This
quadratic equation (\ref{gamma_quadratic}) has a double root at
$\gamma_{\bar f f,IR}=1$, and hence an exact solution of the quadratic
equation (\ref{gamma_quadratic}) yields the same result as the linear
condition (\ref{gamma_eq_1}). However, when applied in the context of
series expansions such as Eq. (\ref{gamma_fund_Deltaseries_fasym}) and
(\ref{gamma_asym_Deltaseries_fasym}), as calculated to finite order,
the results differ from those obtained with the linear condition
(\ref{gamma_eq_1}). This difference arises because the quadratic
condition (\ref{gamma_quadratic_chi}) generates higher-order terms in
powers of the scheme-independent expansion variable, and leads to
different coefficients of lower-order terms \cite{khl,jwlee}.  In a
theory with $N_f$ fermions transforming according to a single
representation of the gauge group, the use of the quadratic condition
(\ref{gamma_quadratic}) was found \cite{khl,jwlee} to (i) show better
convergence as a function of increasing order of truncation of the
series (\ref{gamma_IR_Deltaseries}) than the linear condition
(\ref{gamma_eq_1}) and (ii) predict that the lower boundary ${\cal
  B}_{CW,\ell}$ of the conformal window occurs at a higher value of
$N_f$ than the linear condition.

In a theory with multiple fermions in different representations of the
gauge group, the generalization of the condition (\ref{gamma_eq_1})
for the lower boundary ${\cal B}_{CW,\ell}$ of the conformal window is
that this lower boundary is reached when the larger of the anomalous
dimensions increases through unity, since this would be expected to
result in the dynamical mass generation for the fermion with the
larger anomalous dimension, thereby driving the system out of the
conformal window. Thus, in our present theory, this lower end of the
conformal window occurs if
\beq
    {\rm max}(\gamma_{\bar\psi\psi,IR}, \ \gamma_{\bar\chi\chi,IR}) = 1 \ . 
\label{gamma_condition}
\eeq
In this type of theory, the quadratic form of the critical condition
is Eq. (\ref{gamma_quadratic}) with $\gamma_{\bar f f, IR}$ being given by
${\rm max}(\gamma_{\bar\psi\psi,IR}, \ \gamma_{\bar\chi\chi,IR})$.
Since $\gamma_{\bar\chi\chi,IR} > \gamma_{\bar\psi\psi,IR}$ here,
Eq. (\ref{gamma_quadratic}) reduces to
\beq
\gamma_{\bar \chi \chi,IR}(2-\gamma_{\bar \chi \chi,IR})=1 \ . 
\label{gamma_quadratic_chi}
\eeq
Because of the approximations involved in applying either the linear
condition (\ref{gamma_eq_1}) or the quadratic condition
(\ref{gamma_quadratic}) in the context of finite-order series
expansions, it is useful to compare the lower boundary ${\cal
  B}_{CW,\ell}$ predicted by each of these for the present theory.
The difference gives a measure of the uncertainties involved in the
determination of this lower boundary using the $\gamma$CC
condition. The boundary ${\cal B}_{CW,\ell}$ was calculated in
\cite{khl} using the quadratic $\gamma$CC condition. We have checked
and confirmed the result for ${\cal B}_{CW,\ell}$ obtained in
\cite{khl} with the quadratic $\gamma$CC condition. For the
comparison, here we will calculate the prediction for this boundary
using the linear condition.

As a side note to our study, it may be recalled that the conditions
(\ref{gamma_le_1}) and (\ref{gamma_condition}) have a connection to
approaches to physics beyond the Standard Model involving dynamical
electroweak symmetry breaking (EWSB). In such approaches there has
been interest in models featuring a new gauge interaction that becomes
strongly coupled on the TeV scale, producing fermion condensates and
thus EWSB.  Models with the property of having a slowly running gauge
coupling and approximate scale invariance over an extended interval of
Euclidean energy/momentum scales, due to an approximate zero of the
relevant beta function, have been of particular interest.  One reason
for this is that when the approximate scale invariance in the theory
is dynamically broken by the formation of fermion condensates, this
gives rise to an approximate Nambu-Goldstone boson, namely a dilaton
\cite{dilaton}. In turn, insofar as the observed Higgs boson is
modelled as a composite particle, at least partially dilatonic in
nature, this can provide a means of helping to protect its mass aginst
large radiative corrections.  Although the observed properties of the
Higgs boson, including the production cross section and couplings to
the $W$ and $Z$ vector bosons and to Standard-Model fermions, are in
excellent agreement with SM predictions \cite{higgs_exp,pdg},
experimental work will continue to search for, and set constraints on,
Higgs compositeness and possible deviations from SM predictions. A
second reason is that a renormalization-group flow from the UV to the
IR that is influenced by an approximate IRFP can naturally give rise
to large anomalous dimension(s) $\gamma_{\bar f f, IR} \simeq 1$ for
the fermions $f$ subject to the strongly coupled gauge
interaction. This has been useful in the effort to produce a
realistically large top quark mass while suppressing flavor-changing
neutral-current processes and minimizing corrections to precision
electroweak observables. (In this model-building effort, one must also
confront the challenge of producing the requisite large splitting
between the $t$ and $b$ quark masses.)  Examples of reasonably
UV-complete models with dynamical EWSB that also feature sequential
breaking of an extended gauge symmetry to produce a generational
hierarchy in quark and charged lepton masses, as well as neutrino
masses, and make use of this $\gamma_{\bar f f,IR} \simeq 1$ property,
are discussed, e.g., in \cite{ntckm}.  With fermions in a single
representation of the gauge group, such as SU(3) with $N_F=8$ fermions
in the fundamental representation, lattice simulations
\cite{lsd_2016}-\cite{lsd_2019} have found an anomalous dimension
$\gamma_m \sim 1$ for the strongly coupled fermion and have shown that
the spectrum of the theory includes a light $0^{++}$ state consistent
with being an approximate dilaton. Lattice simulations have also been
carried out for other models, including an SU(3) theory with
two <flavors of fermions in the sextet representation \cite{sextet}.

We recall that a rigorous upper bound on $\gamma_{\bar f f,IR}$ in a
conformal field theory is that \cite{mack,gir,nakayama}
\beq
\gamma_{\bar f f, IR}   \le 2 \ ,
\label{cft_bound}
\eeq
where here, $f$ refers to any fermion in the theory.  This is
evidently less restrictive than the bound (\ref{gamma_le_1}) and need
not be saturated at the lower boundary ${\cal B}_{CW,\ell}$ of the
conformal window.

In passing, it should be mentioned that an approximate condition for
spontaneous chiral symmetry breaking via formation of the condensate
$\langle \bar f f \rangle$ derived from analysis of the
Schwinger-Dyson equation for the $f$ fermion propagator is that this
occurs as the coupling $\alpha$ exceeds the value $\alpha_{cr} =
\pi/(3C_f)$.  As applied to estimate the lower boundary of the
conformal window, this would be a condition on the value of
$\alpha_{IR}$ at the IRFP as one approaches this lower boundary.
While this is a reasonable rough guide, the maximal scheme-independent
level to which it can be applied is the two-loop level, since the
value of the $n$-loop ($n\ell$) IR coupling $\alpha_{IR,n\ell}$ at
higher-loop level, as calculated from the beta function
(\ref{beta_series}), is scheme-dependent. Furthermore, as one
approaches the strongly coupled regime near the lower end of the
conformal window, the value of the IR zero of the $n$-loop beta
function, $\alpha_{IR,n\ell}$, changes substantially as one goes from
two-loop order to higher-loop order.  For example, as listed in Table
II of \cite{gsi}, for SU(3) with $N_F=9$ fermions in the fundamental
representation, $\alpha_{IR,2\ell}=5.24$, while
$\alpha_{IR,3\ell}=1.03$, as calculated in the widely used
$\overline{\rm MS}$ scheme.  Consequently, here we focus on the
$\gamma$CC condition (\ref{gamma_condition}), since it can be applied
in a scheme-independent manner.

% ===================================================================

% section III 
\section{Scheme-Independent Calculation of Anomalous Dimensions of Fermion
  Bilinear Operators}
\label{general_results_section}

In this section, for a theory with an SU($N_c$) gauge group and
(massless) fermion content consisting of $N_F$ fermions in the
fundamental representation, $F$, and $N_{A_2}$ fermions in the
antisymmetric rank-2 representation, $A_2$, we present explicit
calculations of the coefficients $\kappa^{(F)}_j$ and $\kappa^{(A_2)}_j$
with $j=1,2,3$ using the scheme-independent expansions of the
anomalous dimensions $\gamma_{\bar\psi\psi,IR}$ and
$\gamma_{\bar\chi\chi,IR}$ in Eqs. (\ref{gamma_IR_Deltaseries}) and
the analogue for $\gamma_{\bar\chi\chi,IR}$ with $1 \le j \le 3$.
These yield the anomalous dimensions
$\gamma_{\bar\psi\psi,IR}$ and $\gamma_{\bar\chi\chi,IR}$ up to
$O(\Delta_F^3)$ and $O(\Delta_{A_2}^3)$, respectively.

It is convenient to define factors that occurs repeatedly in the 
denominators of $\gamma_{IR,F}$ and $\gamma_{IR,A_2}$, namely 
\beq
{\cal D}_F = N_c(25N_c^2-11) + 2N_{A_2}(N_c-2)(N_c+1)(N_c-3) 
\label{dcal_fund}
\eeq
and
\beq
{\cal D}_{A_2} = N_c(18N_c^2-11N_c-22)-N_F(N_c-3)(N_c+1) \ . 
\label{dcal_asym}
\eeq

For the first two coefficients we calculate 
\beq
\kappa^{(F)}_1 = \frac{4(N_c^2-1)}{{\cal D}_F} \ , 
\label{kappa1_fund}
\eeq
\beq
\kappa^{(A_2)}_1 = \frac{4(N_c-2)^2(N_c+1)}{{\cal D}_{A_2}} \ ,
\label{kappa1_asym}
\eeq
\begin{widetext} 
\beq
\kappa^{(F)}_2 = \frac{4(N_c^2-1)}{3{\cal D}_F^3} \, \Bigg [ 
  N_c(9N_c^2-2)(49N_c^2-44) + 4N_{A_2}(N_c-2)(N_c+1)(N_c-3)(3N_c-2)(5N_c+3)
  \Bigg ] \ ,
\label{kappa2_fund}
\eeq
and
\beq
\kappa^{(A_2)}_2 = \frac{(N_c-2)^3(N_c+1)}{3{\cal D}_{A_2}^3} \, \Bigg [ 
  N_c(11N_c^2-4N_c-8)(93N_c^2-88N_c-176) - 2N_F(N_c-3)(N_c+1)(37N_c^2-16N_c-33)
  \Bigg ] \ .
\label{kappa2_asym}
\eeq
\end{widetext}

Following our notation in \cite{dexm}, we write the third-order coefficients
in the form 
\beq
\kappa^{(F)}_3 = \frac{8(N_c^2-1)}{27{\cal D}_F^5} 
\Bigg [ A^{(F)}_0 + A^{(F)}_1 N_{A_2} + A^{(F)}_2 N_{A_2}^2 
+ A^{(F)}_3 N_{A_2}^3 \Bigg ] 
\label{kappa3_fund}
\eeq
and
\begin{widetext}
\beq
\kappa^{(A_2)}_3 = \frac{(N_c-2)^3(N_c+1)}{54{\cal D}_{A_2}^5} 
\Bigg [ A^{(A_2)}_0 + A^{(A_2)}_1 N_F + A^{(A_2)}_2 N_F^2 
+ A^{(A_2)}_3 N_F^3 \Bigg ] \ , 
\label{kappa3_asym}
\eeq
and we calculate 
\beq
A^{(F)}_0 =
N_c^2\bigg [\Big ( 274243N_c^8-455426N_c^6-114080N_c^4+47344N_c^2+35574 \Big )
- 4224N_c^2(4N_c^2-11)(25N_c^2-11)\zeta_3 \, \bigg ] \ ,
\label{kappa3_fund_part0}
\eeq
\beqs
A^{(F)}_1 &=& 4N_c(N_c-2)(N_c-3)\bigg [ \Big (
16981N_c^7+35460N_c^6+42927N_c^5+47342N_c^4+9432N_c^3-12849N_c^2 \cr\cr
&-&18843N_c-11616 \Big )
-576N_c^2\Big ( 25N_c^4+198N_c^3+187N_c^2-121N_c-121\Big )\zeta_3 \,
\bigg ] \ , 
\label{kappa3_fund_part1}
\eeqs
\beqs
A^{(F)}_2 &=& 8(N_c-2)(N_c-3)\bigg [ \Big (
  689N_c^8-1402N_c^7-9208N_c^6-15693N_c^5-9219N_c^4+16662N_c^3+19860N_c^2
  \cr\cr
  &+&10617N_c+5598\Big ) -192N_c^2\Big (3N_c^5-65N_c^4-238N_c^3-165N_c^2
  +231N_c+198\Big )\zeta_3 \, \bigg ] \ ,
\label{kappa3_fund_part2}
\eeqs
\beq
A^{(F)}_3 = 128N_c(N_c-2)^2(N_c-3)^2(N_c+1)(3N_c^2+7N_c+6)(-11+24\zeta_3) \ ,
\label{kappa3_fund_part3}
\eeq
\beqs
A^{(A_2)}_0 &=&
N_c^2\bigg [ \Big (
  1670571N_c^9-7671402N_c^8+2181584N_c^7+25294256N_c^6-13413856N_c^5
  \cr\cr
 &-&17539136N_c^4+16707328N_c^3+3046912N_c^2-27320832N_c-18213888 \Big ) \cr\cr
  &-&8448N_c^2(N_c+2)(18N_c^2-11N_c-22)(3N_c^3-28N_c^2+176)\zeta_3 \,
  \bigg ] \ ,
\label{kappa3_asym_part0}
\eeqs
\beqs
A^{(A_2)}_1 &=& -4N_c(N_c-3)\bigg [ \Big (
  60552N_c^8-150015N_c^7-373894N_c^6+138737N_c^5+300380N_c^4+421197N_c^3
  +768345N_c^2 \cr\cr
  &+&858660N_c+435468 \Big )
  -192N_c^2\Big ( 141N_c^5-2075N_c^4-6226N_c^3+1056N_c^2+17424N_c+11616
  \Big ) \zeta_3 \, \bigg ] \ ,
\label{kappa3_asym_part1}
\eeqs
\beqs
A^{(A_2)}_2 &=& 8(N_c-3)\bigg [ \Big (
  1148N_c^8-3919N_c^7-17365N_c^6-5724N_c^5+35724N_c^4+84915N_c^3+70641N_c^2
  \cr\cr
  &+&32928N_c+15588 \Big )
  -192N_c^2\Big ( 3N_c^5-164N_c^4-271N_c^3+396N_c^2+1320N_c+792 \Big
  )\zeta_3 \, \bigg ]
\ ,
\label{kappa3_asym_part2}
\eeqs
and
\beq
A^{(A_2)}_3 =-128N_c(N_c+1)(N_c-3)^2(3N_c^2+7N_c+6)(-11+24\zeta_3) \ .
\label{kappa3_asym_part3}
\eeq
\end{widetext}
Here, $\zeta_s = \sum_{n=1}^\infty n^{-s}$ is the Riemann zeta
function, and $\zeta_3 = 1.2020569...$ We have remarked above on the
reason for the occurrence of $(N_c-2)$ factors (or powers thereof) in
conjunction with the numbers $N_{A_2}$. The occurrence of $(N_c-3)$ factors
in various expressions reflects the reduction of the theory to one with
$N_F+N_{A_2}$ fermions in the fundamental representation in the case $N_c=3$.
We straightforwardly check that our general-$N_c$ results above satisfy
the identities (\ref{su3_identities}). 

% ================================================================

\section{SU(4) Theory}
\label{su4_section}

In this section, for the case $N_c=4$, i.e.,
$G={\rm SU}(4)$, we list the special cases of
the general-$N_c$ expressions for the coefficients $\kappa^{(F)}_j$
and $\kappa^{(A_2)}_j$ with $1 \le j \le 3$ and the resultant
expressions for $\gamma_{\bar\psi\psi,IR,\Delta_F^p}$ and
$\gamma_{\bar\chi\chi,IR,\Delta_{A_2}^p}$ with $1 \le p \le 3$.

% ==============================================

\subsection{Direct Calculations of Anomalous Dimensions}

For $N_c=4$, the upper end of the conformal window, defined by the
condition $b_1 > 0$, is
\beq
N_F+2N_{A_2} < 22 \ ,
\label{b1pos_su4}
\eeq
and the condition that the two-loop beta function should have an IR zero is
\beq
205N_F+440N_{A_2} > 2176 \ .
\label{b2neg_su4}
\eeq
As before, we denote the region in the first quadrant of the
$(N_F,N_{A_2})$ plane where the inequalities (\ref{b1pos_su4}) and
(\ref{b2neg_su4}) are simultaneously satisfied as $I_{IRZ}$. The lines
that are the upper and lower boundaries of the $I_{IRZ}$ region have
slopes that are almost equal. From Eq. (\ref{b1zero_line_slope}), the
upper boundary, i.e., the solution to the condition $b_1=0$ has slope
$dN_{A_2}/dN_F = -1/2$, while from Eq. (\ref{b2zero_line_slope}), the
lower $I_{IRZ}$ boundary, i.e., the solution to the condition $b_2=0$,
has slope $dN_{A_2}/dN_F = -41/88 = -0.4659$.  Regarding the figures
to be presented below, we note that if one sets $N_F=4$, then the
$I_{IRZ}$ interval in $N_{A_2}$ is $3.082 < N_{A_2} < 9$ and if one
sets $N_{A_2}=4$, then the $I_{IRZ}$ interval in $N_F$ is $2.029 < N_F <
14$. In the (4,4,4) theory, $\Delta_F=10$ and $\Delta_{A_2}=5$. 

Substituting $N_c=4$ in the results for $\kappa^{(F)}_j$ and
$\kappa^{(A_2)}_j$ given for $G={\rm SU}(N_c)$ in the previous section,
we find the following: 
\beq
\kappa^{(F)}_1 = \frac{15}{389+5N_{A_2}} \ ,
\label{kappa1_fund_su4}
\eeq
\smallskip

\beq
\kappa^{(F)}_2 = \frac{25(5254+115N_{A_2})}{(389+5N_{A_2})^3} \ , 
\label{kappa2_fund_su4}
\eeq
\beq
\kappa^{(A_2)}_1 = \frac{80}{888-5N_F} \ , 
\label{kappa1_asym_su4}
\eeq
and
\beq
\kappa^{(A_2)}_2 = \frac{400(19456-165N_F)}{(888-5N_F)^3} \ .
\label{kappa2_asym_su4}
\eeq
\begin{widetext}
\beqs
\kappa^{(F)}_3 &=& \frac{5}{36(389+5N_{A_2})^5} \, \bigg [
  (8039476475-696689664\zeta_3) +(479848740-197766144\zeta_3)N_{A_2} \cr\cr
  &+&(-16264767+46568448\zeta_3)N_{A_2}^2
  +(-288640+629760\zeta_3)N_{A_2}^3 \, \bigg ] \ ,
\label{kappa3_fund_su4}
\eeqs
and
\beqs
\kappa^{(A_2)}_3 &=& \frac{640}{27(888-5N_F)^5} \, \bigg [
    (28645111296+7201751040\zeta_3)-(120552246+1055342592\zeta_3)N_F \cr\cr
    &+&(-12526131+33675264\zeta_3)N_F^2 + (72160-157440\zeta_3)N_F^3 \, \bigg ]
\ .
\label{kappa3_asym_su4}
\eeqs
\end{widetext}

For the theory with $N_c=4$, i.e., $G={\rm SU}(4)$, $N_F=4$, and
$N_{A_2}=4$, our general expressions yield the following
(where floating-point values are quoted to the indicated precision):
\beq
\kappa^{(F)}_1 = \frac{15}{409} = 3.6675 \times 10^{-2} \ ,
\label{kappa1_fund_444}
\eeq
\beq
\kappa^{(F)}_2 = \frac{142850}{(409)^3} = 2.0879 \times 10^{-3} \ ,
\label{kappa2_fund_444}
\eeq
\beqs
\kappa^{(F)}_3 &=& \frac{48400811015}{36 \cdot (409)^5} -
\frac{715520}{3 \cdot (409)^4} \zeta_3 \cr\cr
&=& 2.37475 \times 10^{-4} \ ,
\label{kappa3_fund_444}
\eeqs
\beq
\kappa^{(A_2)}_1 = \frac{20}{217} = 0.092166 \ ,
\label{kappa1_asym_444}
\eeq
\beq
\kappa^{(A_2)}_2 = \frac{117475}{(217)^3} = 1.1497 \times 10^{-2} \ ,
\label{kappa2_asym_444}
\eeq
and
\beqs
\kappa^{(A_2)}_3 &=& \frac{17479439035}{27 \cdot (217)^5} +
\frac{3368960}{9 \cdot (217)^4} \zeta_3 \cr\cr
&=& 1.5484 \times 10^{-3} \ ,
\label{kappa3_asym_444}
\eeqs
where partial factorizations are shown for denominators. 

Substituting these coefficients into
Eqs. (\ref{gamma_fund_Deltaseries_fasym}) and
     (\ref{gamma_asym_Deltaseries_fasym}),
with $\Delta_F=2\Delta_{A_2}=10$ for this (4,4,4) theory, we have 
\beq
\gamma_{\bar\psi\psi,IR,\Delta_F} = 0.367 \ ,
\label{gamma_fund_p1_444}
\eeq
\beq
\gamma_{\bar\psi\psi,IR,\Delta_F^2} = 0.576 \ ,
\label{gamma_fund_p2_444}
\eeq
\beq
\gamma_{\bar\psi\psi,IR,\Delta_F^3} = 0.683 \ ,
\label{gamma_fund_p3_444}
\eeq
\beq
\gamma_{\bar\chi\chi,IR,\Delta_{A_2}} = 0.461 \ ,
\label{gamma_asym_p1_444}
\eeq
\beq
\gamma_{\bar\chi\chi,IR,\Delta_{A_2}^2} = 0.748 \ ,
\label{gamma_asym_p2_444}
\eeq
and
\beq
\gamma_{\bar\chi\chi,IR,\Delta_{A_2}^3} = 0.942 \ .
\label{gamma_asym_p3_444}
\eeq

Because $\kappa^{(F)}_j$ and $\kappa^{(A_2)}_j$ are positive for all
of the orders $j=1,2,3$ for which we have calculated them, several
monotonicity relations follow.  These are the analogues, in the
current theory, of the relations noted in \cite{dexm}. First, for
these orders, with fixed $\Delta_F=2\Delta_{A_2}$, the anomalous dimensions
$\gamma_{\bar\psi\psi,IR,\Delta_F^p}$ and
$\gamma_{\bar\chi\chi,IR,\Delta_{A_2}^p}$ are monotonically increasing
functions of $p$. Second, for a fixed $p$,
$\gamma_{\bar\psi\psi,IR,\Delta_F^p}$ is a monotonically increasing
function of $\Delta_F$ and $\gamma_{\bar\chi\chi,IR,\Delta_{A_2}^p}$
is a monotonically increasing function of $\Delta_{A_2}$. 

Since finite-order perturbative calculations of this type tend to
become progressively less accurate as one approaches the lower
boundary ${\cal B}_{CW,\ell}$ of the conformal window, one is
motivated to assess the effect of higher-order corrections.  One
approach for this purpose is to perform a rough extrapolation (ext) of
our results for $p=1,2,3$ to $p=\infty$.  This yields values for
$\gamma_{\bar\psi\psi,IR}$ and $\gamma_{\bar\chi\chi,IR}$ that we
estimate to be approximately 10-20 \% larger than our respective
$\gamma_{\bar\psi\psi,IR,\Delta_F^3}$ and
$\gamma_{\bar\chi\chi,IR,\Delta_F^3}$ values, namely
$\gamma_{\bar\psi\psi,IR,{\rm ext}} \simeq 0.7 - 0.8$ and
$\gamma_{\bar\chi\chi,IR,{\rm ext}} \simeq 1.0 - 1.1$. We next compare
these results with the values obtained from lattice simulations in
Ref. \cite{fx444}, namely $\gamma^{(4)}_m \simeq 0.75$ and
$\gamma^{(6)}_m \simeq 1.0$. (Recall the equivalences of notation for
this SU(4) theory: $\gamma^{(4)}_m \equiv \gamma_{\bar\psi\psi,IR}$
and $\gamma^{(6)} \equiv \gamma_{\bar\chi \chi,IR}$.)  To within the
uncertainties in our extrapolation and in the lattice measurements,
our calculations are in agreement with these values of anomalous
dimensions obtained in \cite{fx444}.  This agreement between our
perturbative calculations, which require an IR fixed point in the
conformal window, and the values of these anomalous dimensions
measured in the lattice simulations, is consistent with the conclusion
in \cite{fx444} that this theory is in the conformal window (near the
lower boundary, since the measured $\gamma^{(6)}_m \simeq 1$ and our
$\gamma_{\bar\chi\chi,IR,{\rm ext}} \simeq 1$).  A cautionary remark
is that the uncertainties in the perturbative calculation of anomalous
dimensions are substantial at the lower end of the conformal window,
as are the uncertainties in our rough extrapolation.

% =================================================

\subsection{Calculation of Pad\'e Approximants for Anomalous Dimensions}

Another useful approach to estimating anomalous dimensions from finite
series expansions is the use of Pad\'e approximants, and we have used
these in our earlier work for theories with fermions in a single
representation \cite{flir,gsi,dex,dexl,pgb}.  Given a series expansion
calculated to a finite order, a $[p,q]$ Pad\'e approximant is a
rational function with numerator and denominator having respective
degrees $p$ and $q$ in the expansion variable and satisfying the
property that the Taylor series expansion of this rational function
fits all of the coefficients in the original series expansion
\cite{pade_review,padenotation}.  In our present context, for a given
fermion $f$ (equal to $\psi$ in the $F$ representation or $\chi$ in
the $A_2$ representation of SU4)), let us consider the
scheme-independent expansion calculated to order $s$:
\beqs
\gamma_{\bar f f,IR,\Delta_f^s} &=&
\sum_{j=1}^s \kappa^{(f)}_j \Delta_f^j \cr\cr
&=&  \kappa^{(f)}_1 \Delta_f \Big [
  1 + \frac{1}{\kappa^{(f)}_1} \sum_{j=2}^s \kappa^{(f)}_j \Delta_f^{j-1}
  \Big ] \ . \cr\cr
&&
\label{gamma_reduced}
\eeqs
We calculate the $[p,q]$ Pad\'e approximant to the expression in square
brackets, which has the form 
\beq
\gamma_{\bar f f,IR,[p,q]} =
\kappa^{(f)}_1 \Delta_f \bigg [
  \frac{1 + \sum_{i=1}^p {\cal N}_{f,i}\Delta_f^i}
{1 + \sum_{j=1}^q D_{f,j} \Delta_f^j} \bigg ] \ ,
\label{gamma_pqpade}
\eeq
where $p+q=s-1$. With $s=3$, the possible Pad\'e approximants to the
expression in square brackets are then [2,0], [1,1], and [0,2].  The
[2,0] approximant is just the original series, which we have already
used to calculate $\gamma_{\bar f f, IR,\Delta_f^3}$ for $f=\psi$ and
$f=\chi$, so we focus on the [1,1] and [0,2] approximants here.  In
addition to providing a closed-form rational-function approximation to
the finite series (\ref{gamma_reduced}), a Pad\'e approximant also can
be used in another way, namely to yield an estimate of the effects of
higher-order terms. 

By construction, the $[p,q]$ Pad\'e approximant in
(\ref{gamma_pqpade}) is analytic at $\Delta_f=0$, and if it has $q >
0$, then it is a meromorphic function with $q$ poles.  The radius of
convergence of the Taylor series expansion of the $[p,q]$ Pad\'e
approximant is set by the magnitude of the pole nearest to the origin
in the complex $\Delta_f$ plane.  Consequently, a necessary condition
that must be satisfied for a Pad\'e approximant to be useful for our
analysis here is that, considered as a function of the general
variable $\Delta_f$, it should not have a pole at any $\Delta_{f,{\rm
    pole}}$ that is closer to the origin than the actual value of
$\Delta_f$ in the theory of interest. We recall that for the (4,4,4)
theory, $\Delta_F=10$ and $\Delta_{A_2}=5$. A further caveat with this
method is that if a $[p,q]$ Pad\'e approximant has a pole that is near
to the physical value of the expansion parameter, even if it is
farther from the origin, this might produce a spuriously large value
of the anomalous dimension.

We focus here on approximants for $\gamma_{\bar\chi\chi,IR}$, since it is
larger than $\gamma_{\bar\psi\psi,IR}$. 
We calculate the following [1,1] and [0,2] Pad\'e approximants for
this anomalous dimension: 

\beq
\gamma_{\bar\chi\chi,IR,[1,1]} = 0.092166\Delta_{A_2}
\Bigg [ \frac{1 - 0.0099444\Delta_{A_2}}{1-0.13468 \Delta_{A_2}} \Bigg ] \ , 
\label{gamma_asym_fx444_pade11}
\eeq
and
\begin{widetext}
\beq
\gamma_{\bar\chi\chi,IR,[0,2]} = 0.092166\Delta_{A_2}
\Bigg [ \frac{1}{1-0.12474\Delta_{A_2} - 0.0012404\Delta_{A_2}^2} \Bigg ] \ .
\label{gamma_asym_fx444_pade02}
\eeq
\end{widetext}
The approximant
$\gamma_{\bar\chi\chi,IR,[1,1]}$ has a pole at $\Delta_{A_2}=7.42492$, and 
$\gamma_{\bar\chi\chi,IR,[0,2]}$ has poles at $\Delta_{A_2}=7.46299$ and
$-108.022$. These are farther from the origin than the physical value
$\Delta_{A_2}=5$, although the poles at 7.42 and 7.46 are moderately close
to the physical value, $\Delta_{A_2}=5$. Evaluating these 
approximants at this value of $\Delta_{A_2}$, we obtain
$\gamma_{\bar\chi\chi,IR,[1,1]} = 1.34$ and 
$\gamma_{\bar\chi\chi,IR,[0,2]} = 1.33$, somewhat larger than our
rough extrapolations discussed above.  

% =================================================

\subsection{Estimates of Lower Boundary of Conformal Window} 

Ref. \cite{fx444} observed that its conclusion that the
$(N_c,N_F,N_{A_2})=(4,4,4)$ theory has an IRFP, and is thus in the
conformal window, disagreed with the lower boundary ${\cal
  B}_{CW,\ell}$ of the conformal window presented in \cite{khl} on the
basis of the $\gamma$CC condition in quadratic form
(\ref{gamma_quadratic}).  As noted in \cite{fx444}, the theory
(4,4,4) theory is below the lower boundary of the conformal window
from the $\gamma$CC condition shown in Fig. 4 of \cite{khl} as a
function of $(N_F,N_{A_2})$ and reproduced in Fig. 1 of \cite{fx444}.
(In referring to Fig. 4 in \cite{khl}, we remind the reader that the
symbol $N_{A_2}$ used in that figure is the number of Majorana $A_2$
fermions and hence is equal to $2N_{A_2}$ in our notation, where our
$N_{A_2}$ is the number of Dirac $A_2$ fermions.) So the implication
from the lower boundary ${\cal B}_{CW,\ell}$ in \cite{khl} is that the
(4,4,4) theory is in the chirally broken phase, not in the conformal
window. 

To investigate this further, we have performed an alternative
calculation of ${\cal B}_{CW,\ell}$ using our results for
$\gamma_{\bar\psi\psi,IR,\Delta_F^3}$ and
$\gamma_{\bar\chi\chi,IR,\Delta_F^3}$ in conjunction with 
the linear $\gamma$CC critical condition, 
\beq
{\rm max}(\gamma_{\bar\psi\psi,IR,\Delta_F^3}, \
\gamma_{\bar\chi\chi,IR,\Delta_{A_2}^3}) = 1 \ .
\label{gammacc_p3_gen}
\eeq
In applying this condition, the maximal anomalous dimension is
$\gamma_{\bar\chi\chi,IR,\Delta_{A_2}^3}$, which is larger here, for a
given $(N_F,N_{A_2})$, than $\gamma_{\bar\psi\psi,IR,\Delta_{A_2}^3}$.
Therefore, Eq. (\ref{gammacc_p3_gen}) reduces to 
\beq
\gamma_{\bar\chi\chi,IR,\Delta_{A_2}^3} = 1 \ .
\label{gammacc_p3}
\eeq

\begin{figure}
  \begin{center}
    \includegraphics[height=8cm]{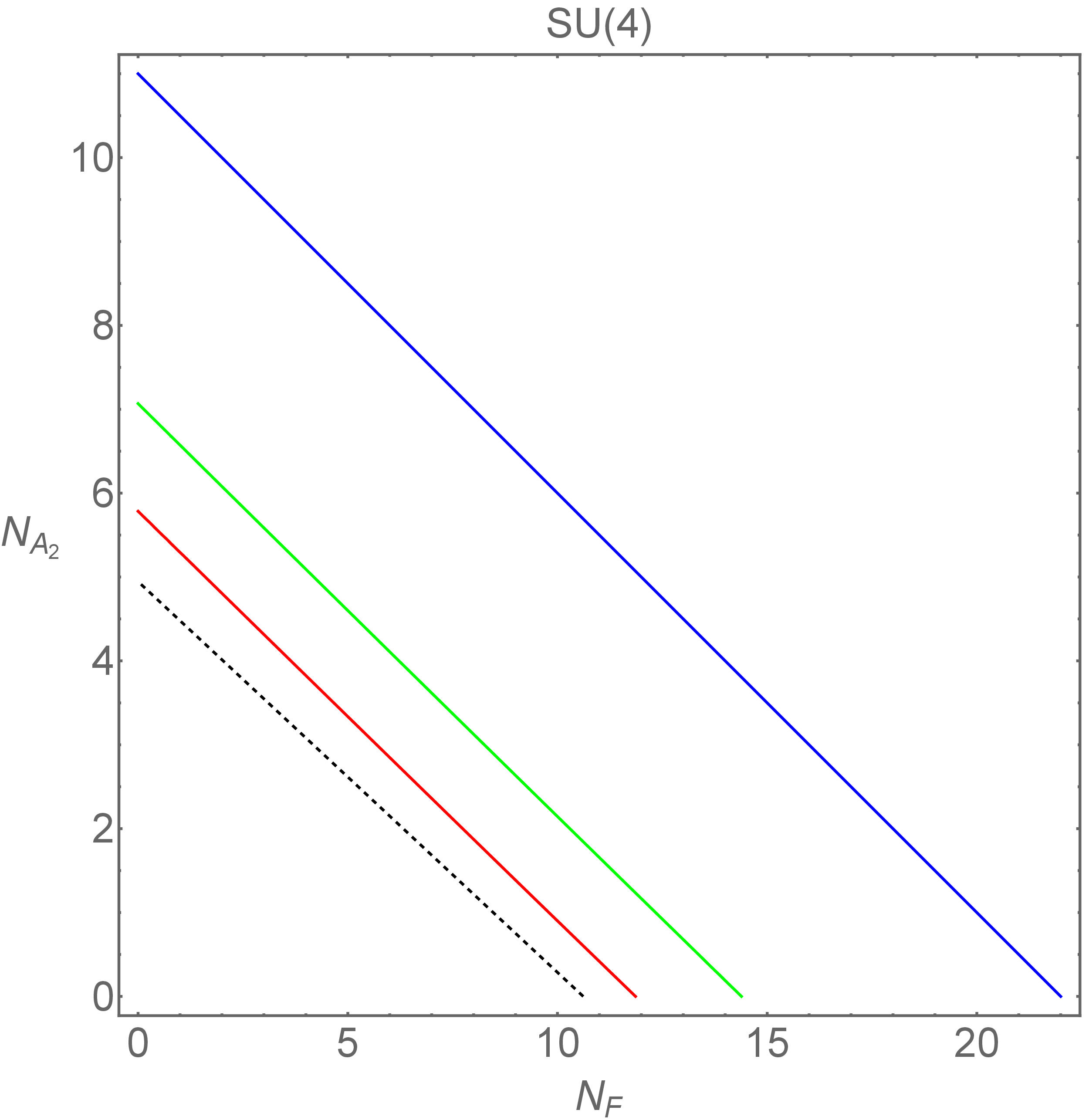}
  \end{center}
  \caption{Plot of regions and boundaries in the $(N_F,N_{A_2})$ plane
    for $G={\rm SU}(4)$. The upper solid line (colored blue) is the
    solution locus of the equation $b_1=0$ and is the upper boundary
    ${\cal B}_{CW,u}$ of the conformal window. The plot shows the
    locations of the boundary ${\cal B}_{CW,\ell}$, as calculated in
    \cite{khl} from the quadratic $\gamma$CC condition
    (colored green), and as calculated here from the linear
    $\gamma$CC condition (colored red).
    The dashed line is the solution locus of the
    equation $b_2=0$ and is the lower boundary ${\cal B}_{IRZ,\ell}$
    of the IRZ region.}
    \label{region_plot}
\end{figure}

We show our results in Fig. \ref{region_plot}.  The uppermost line
(colored blue) is the upper boundary ${\cal B}_{CW,u}={\cal
  B}_{IRZ,u}$ of the conformal window, given by the condition $b_1=0$.
The locations of the lower boundary ${\cal B}_{CW,\ell}$
as calculated in \cite{khl} from the quadratic $\gamma$CC condition
(colored green), and as calculated here from the linear $\gamma$CC
condition (\ref{gammacc_p3_gen}), which reduces to (\ref{gammacc_p3})
(colored red), are shown.  Both of these calculations of ${\cal
  B}_{CW,\ell}$ use the $\kappa^{(F)}_j$ and $\kappa^{(A_2)}_j$
calculated to order $j=3$ from the general results in \cite{dexm}. The
dashed line is the solution locus of the equation $b_2=0$ and is the
lower boundary ${\cal B}_{IRZ,\ell}$ of the IRZ region.  For general
$N_c$ and, in particular, for $N_c=4$, the conditions
(\ref{gammacc_p3_gen}) and (\ref{gammacc_p3}) are nonlinear equations
in the variables $N_F$ and $N_{A_2}$, but the coefficients of the
nonlinear terms are small compared to the coefficients of the linear
terms and get smaller as the total degree $s+t$ of a term $N_F^s
N_{A_2}^t$ increases, so that the solution locus is close to being
linear in the $(N_F,N_{A_2})$ plane. As is evident in
Fig. \ref{region_plot}, the use of the linear form of the $\gamma$CC
in Eq. (\ref{gammacc_p3}) yields a boundary ${\cal B}_{CW,\ell}$ that
lies to the lower left of the boundary ${\cal B}_{CW,\ell}$ obtained
with the use of the quadratic $\gamma$CC condition in the
$(N_F,N_{A_2})$ plane. As was discussed in Sect. \ref{gamma_crit},
this is a consequence of the fact that the quadratic $\gamma$CC
condition (\ref{gamma_quadratic}) generates higher-order terms in
powers of the scheme-independent expansion variables and leads to
different coefficients for lower-order terms.  With ${\cal
  B}_{CW,\ell}$ as determined from (\ref{gammacc_p3}) and shown in
Fig. \ref{region_plot}, the (4,4,4) theory is within the conformal
window, close to the lower boundary.  Along the diagonal
$N_F=N_{A_2}$, the boundary ${\cal B}_{CW,\ell}$ calculated from
(\ref{gammacc_p3}) crosses the point $(N_F,N_{A_2})=(3.88,3.88)$,
slightly to the lower left of the point
$(N_F,N_{A_2})=(4,4)$. Therefore, with ${\cal B}_{CW,\ell}$ computed
via the linear form of the $\gamma$CC condition,
(\ref{gammacc_p3_gen}) or (\ref{gammacc_p3}), the (4,4,4) theory is in
the conformal window.  This is in accord with our result that at cubic
order in the scheme-independent expansion coefficients, the values of
anomalous dimensions that we obtain, namely
$\gamma_{\bar\chi\chi,IR,\Delta_{A_2}^3}=0.942$ in
Eq. (\ref{gamma_asym_p3_444}) and
$\gamma_{\bar\psi\psi,IR,\Delta_F^3}=0.683$ in
Eq. (\ref{gamma_fund_p3_444}) in the (4,4,4) theory, are both less
than 1.  Our comparative analysis showing
the difference in the location of the boundary ${\cal B}_{CW,\ell}$ as
computed via the quadratic $\gamma$CC condition in \cite{khl} and as
computed via the linear $\gamma$CC condition here (with inputs for the
$\kappa^{(A_2)}_j$ and $\kappa^{(F)}_j$ calculated up to the same
maximal order, $j=3$) provides a quantitative measure of the
importance of higher-order terms in the scheme-independent expansions
and hence the uncertainty in the determination of the location of
${\cal B}_{CW,\ell}$.  This comparison makes it clear that these
higher-order corrections are significant. 

Since the anomalous dimensions increase as one moves downward within
the conformal window toward the lower boundary ${\cal B}_{CW,\ell}$,
the linear form of the $\gamma$CC condition implies that for any
theory below this lower boundary, at least some fermion $f$ has an
anomalous dimension $\gamma_{\bar f f, IR}$ that is larger than 1,
where here, $\{ f \} = \{\psi, \ \chi\}$, i.e., $\{ F, A_2 \}$.  A
peculiar feature of the quadratic form of the $\gamma$CC condition is
that if one uses it to determine ${\cal B}_{CW,\ell}$ with input
coefficients $\kappa^{(f)}_j$ calculated to the same maximal order as
with the linear $\gamma$CC condition, then this boundary ${\cal
  B}_{CW,\ell}$ from the quadratic $\gamma$CC condition has the
property that there are theories that lie outside the conformal window
but in which all fermions $f$ have anomalous dimensions $\gamma_{\bar
  f f,IR}$ that are less than 1.  This situation occurs here; our
direct calculation of $\gamma_{\bar\psi\psi,IR}$ and
$\gamma_{\bar\chi\chi,IR}$ in the (4,4,4) theory, with input values
for the $\kappa^{(F)}_j$ and $\kappa^{(A_2)}_j$ computed to the $j=3$
order, yields values for $\gamma_{\bar\chi\chi,IR,\Delta_{A_2}^3}$ and
$\gamma_{\bar\psi\psi,IR,\Delta_F^3}$ that are both less than 1, but
the point $(N_F,N_{A_2})=(4,4)$ lies outside the conformal window, as
calculated in \cite{khl} via the quadratic $\gamma$CC condition with
the same inputs for $\kappa^{(F)}_j$ and $\kappa^{(A_2)}_j$ computed
up to order $j=3$.

\begin{figure}
  \begin{center}
    \includegraphics[height=6cm]{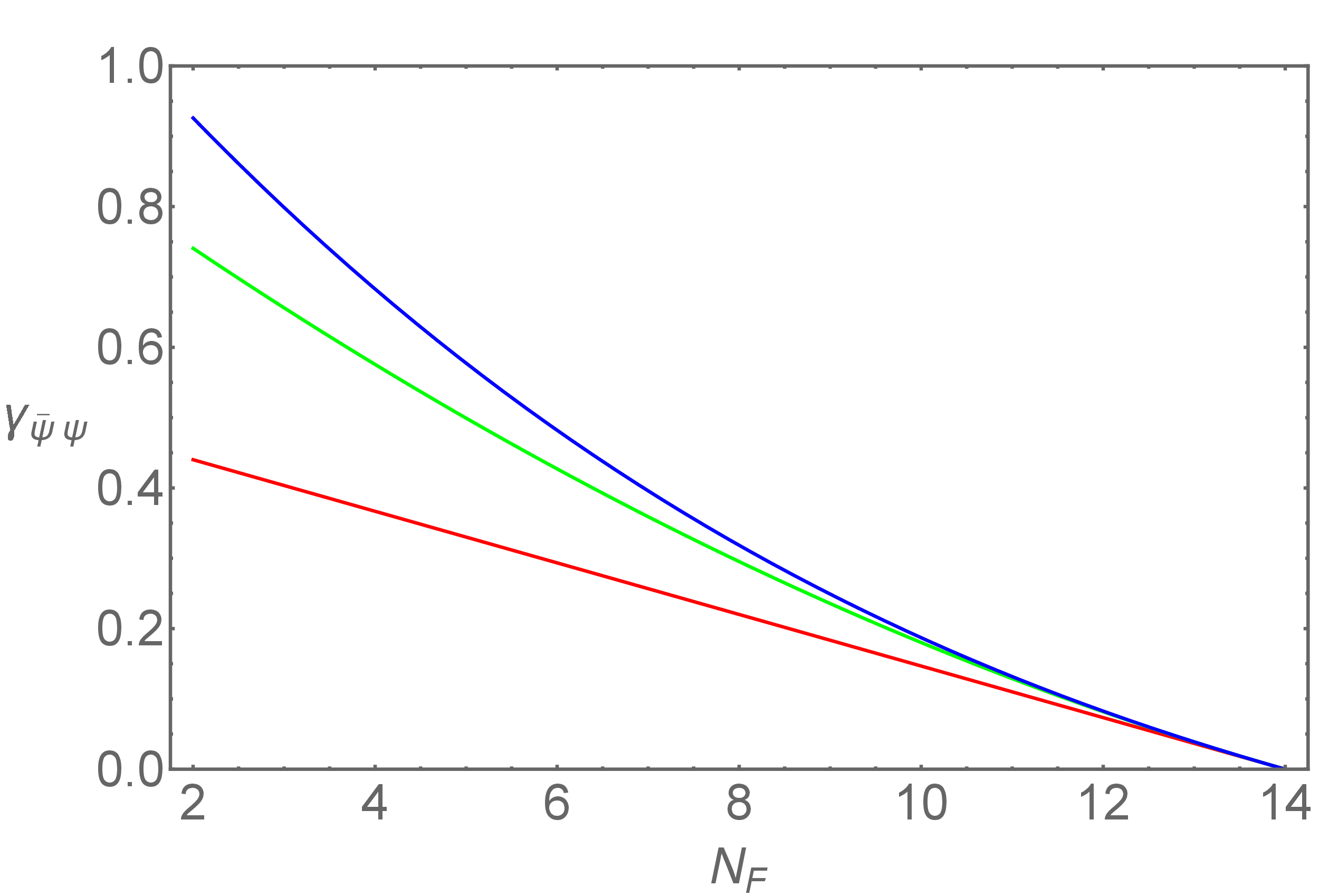}
  \end{center}
  \caption{Plot of $\gamma_{\bar\psi\psi,IR,\Delta_F^p}$ calculated to
    order $p=1,2,3$ for $G={\rm SU}(4)$, and $N_{A_2}=4$, as a
    function of $N_F \in I_{IRZ}$. From bottom to top,
the curves refer to
$\gamma_{\bar\psi\psi,IR,\Delta_F}$ (red),
$\gamma_{\bar\psi\psi,IR,\Delta_F^2}$ (green), and
$\gamma_{\bar\psi\psi,IR,\Delta_F^3}$ (blue). The vertical axis is
simply labelled as $\gamma_{\bar\psi\psi}$ for short.}
\label{gamma_F_NF_plot}
\end{figure}

\begin{figure}
  \begin{center}
    \includegraphics[height=6cm]{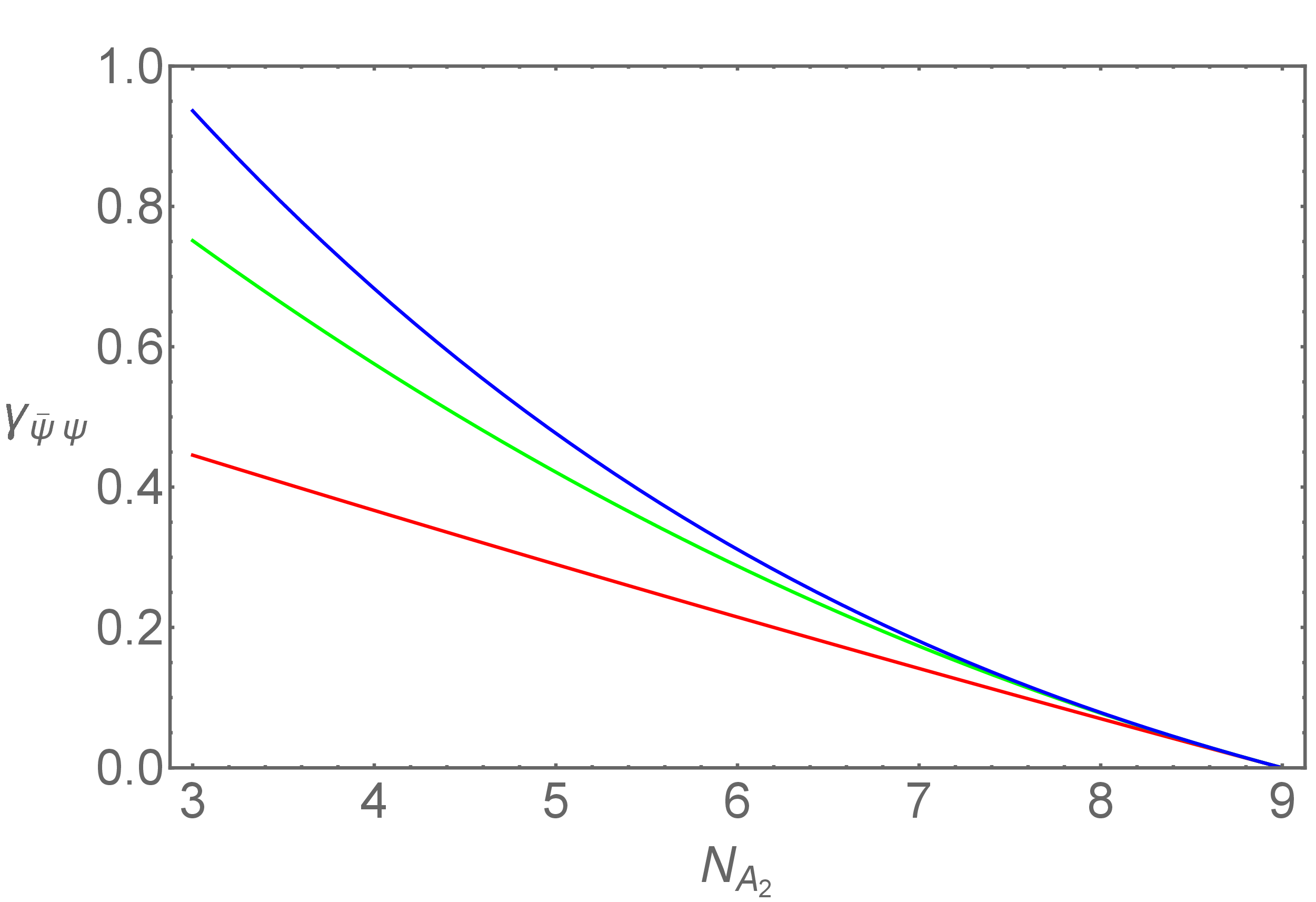}
  \end{center}
  \caption{Plot of $\gamma_{\bar\psi\psi,IR,\Delta_F^p}$ calculated to
    order $p=1,2,3$ for $G={\rm SU}(4)$, and $N_F=4$, as a
    function of $N_{A_2} \in I_{IRZ}$. From bottom to top,
the curves refer to
$\gamma_{\bar\psi\psi,IR,\Delta_F}$ (red),
$\gamma_{\bar\psi\psi,IR,\Delta_F^2}$ (green), and
$\gamma_{\bar\psi\psi,IR,\Delta_F^3}$ (blue).}
\label{gamma_F_NA2_plot}
\end{figure}

\begin{figure}
  \begin{center}
    \includegraphics[height=6cm]{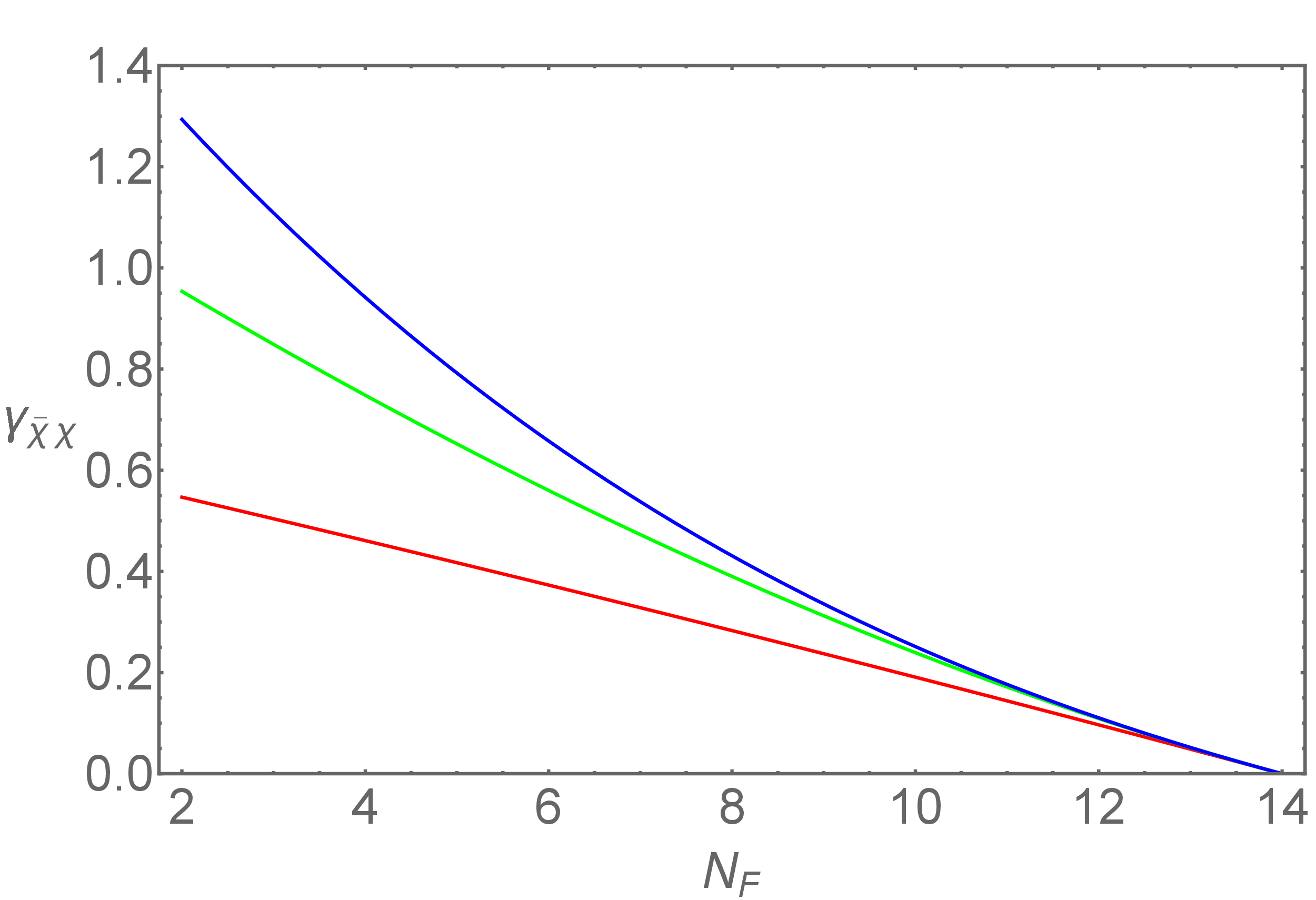}
  \end{center}
  \caption{Plot of $\gamma_{\bar\chi\chi,IR,\Delta_{A_2}^p}$, calculated to
    order $p=1,2,3$ for $G={\rm SU}(4)$, and $N_{A_2}=4$, as a
    function of $N_F \in I_{IRZ}$. From bottom to top,
the curves refer to
$\gamma_{\bar\chi\chi,IR,\Delta_{A_2}}$ (red),
$\gamma_{\bar\chi\chi,IR,\Delta_{A_2}^2}$ (green), and
$\gamma_{\bar\chi\chi,IR,\Delta_{A_2}^3}$ (blue). The vertical axis is
simply labelled as $\gamma_{\bar\chi\chi}$ for short.}
\label{gamma_A2_NF_plot}
\end{figure}

\begin{figure}
  \begin{center}
    \includegraphics[height=6cm]{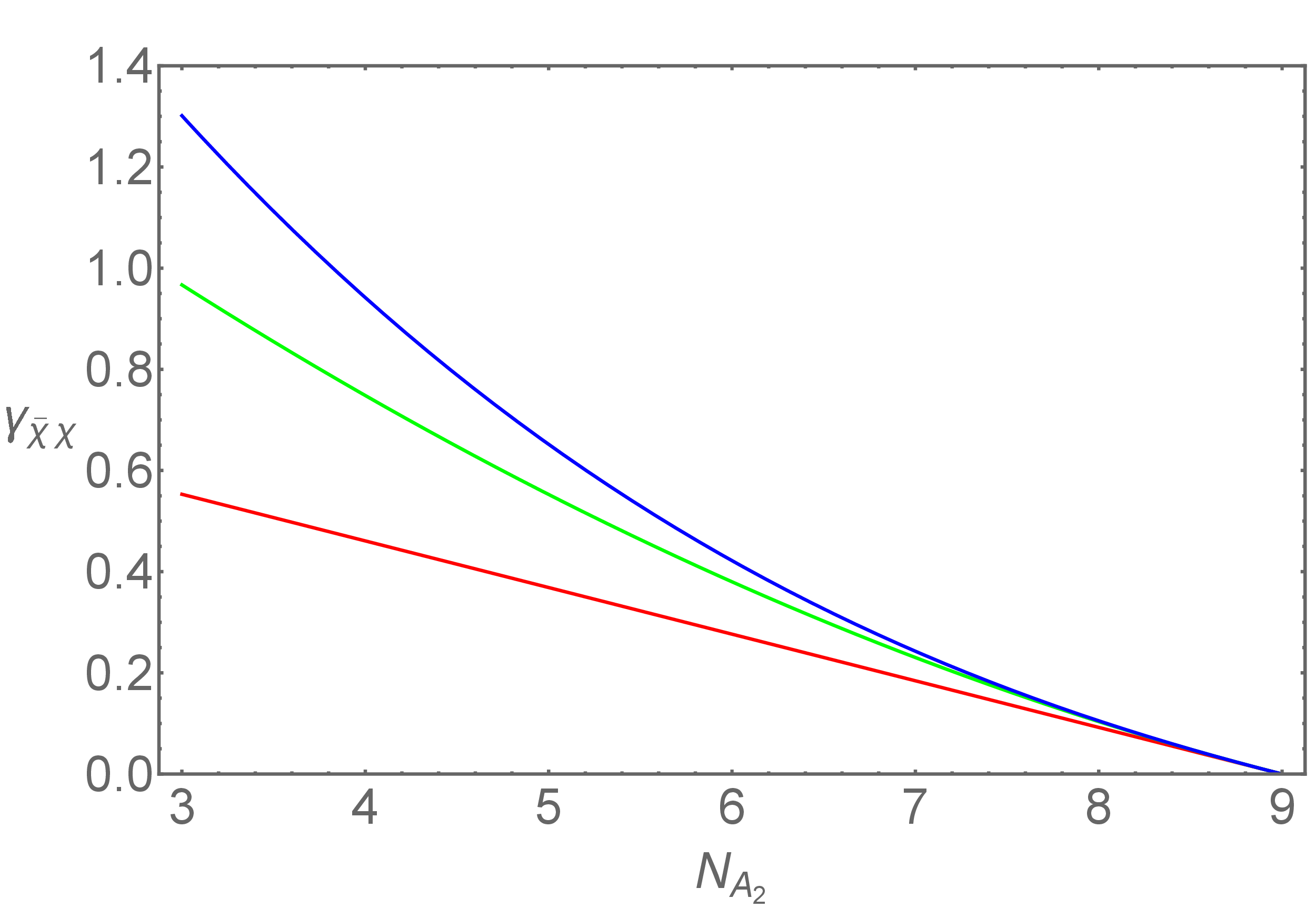}
  \end{center}
  \caption{Plot of $\gamma_{\bar\chi\chi,IR,\Delta_{A_2}^p}$,
    calculated to order $p=1,2,3$ for $G={\rm SU}(4)$, and $N_F=4$, as
    a function of $N_F \in I_{IRZ}$. From bottom to top, the curves
    refer to $\gamma_{\bar\chi\chi,IR,\Delta_{A_2}}$
    (red),$\gamma_{\bar\chi\chi,IR,\Delta_{A_2}^2}$ (green), and
    $\gamma_{\bar\chi\chi,IR,\Delta_{A_2}^3}$ (blue).}
\label{gamma_A2_NA2_plot}
\end{figure}

To investigate the behavior of $\gamma_{\bar\psi\psi,IR,\Delta_F^p}$
and $\gamma_{\bar\chi\chi,IR,\Delta_F^p}$ for $p=1,2,3$ further in this
SU(4) theory, we calculate how they vary as functions of $N_F$ and
$N_{A_2}$, in particular, when one sets $N_F=4$ and varies $N_{A_2}$
or one sets $N_{A_2}=4$ and varies $N_F$. These intervals are,
respectively, a vertical and a horizontal line segment in the
$(N_F,N_{A_2})$ plane, which both pass through the point of primary
interest, $(N_F,N_{A_2})=(4,4)$. Our results are presented in
Figs. \ref{gamma_F_NF_plot}-\ref{gamma_A2_NA2_plot}.
As one can see from Figs. \ref{gamma_A2_NF_plot} and
\ref{gamma_A2_NA2_plot}, for $N_F=4$, this yields the value
$N_{A_2} = 3.8$ as being on ${\cal B}_{CW,\ell}$ calculated from
the linear $\gamma$CC condition (\ref{gammacc_p3}), and for
$N_{A_2}=4$, it yields the value
$N_F=3.6$ as being on this boundary.  These calculations thus serve as
a check on our calculation of the boundary ${\cal B}_{CW,\ell}$ from
the linear $\gamma$CC condition (\ref{gammacc_p3}), since one can
verify that this boundary does pass through the points
$(N_F,A_{A_2})=(4.0,3.8)$ and (3.6,4.0). 

% ==================================

% ===================================

% ==================================

% ==================================

% ==================================

One of the interesting features of this SU(4) theory is that the
gauge-singlet particle spectrum contains composite fermion(s),
$\{f_s\}$.  The lattice
simulations in \cite{fx444} yield anomalous dimensions for several
composite-fermion operators, which are found to be $\lsim 0.5$,
smaller than desired for models of a partially composite top
quark. Comparison is made with one-loop perturbative calculations of
the anomalous dimensions for these gauge-singlet composite fermion
operators. In future work, it could be useful to carry out
higher-order scheme-independent perturbative calculations of the
anomalous dimensions for these composite-fermion operators. This is
beyond the scope of our present work, since the requisite higher-order
coefficients $c^{(f_s)}_\ell$ in the conventional series expansions
(\ref{gamma_series}) have not, to our knowledge, been calculated.

% ===============================================================
% ======================================================================

\section{Conclusions}
\label{conclusions_section}

In this paper we have used our general results in \cite{dexm} to
calculate scheme-independent expansions for anomalous dimensions
$\gamma_{\bar\psi\psi,IR}$ and $\gamma_{\bar\chi\chi,IR}$ of the
fermion bilinear operators $\bar\psi\psi$ and $\bar\chi\chi$ at an
infrared fixed point in an asymptotically free SU($N_c$) gauge theory
with massless fermion content consisting of $N_F$ fermions $\psi^a_i$
in the fundamental representation and $N_{A_2}$ fermions $\chi^{ab}_j$
in the antisymmetric rank-2 tensor representation. These calculations
were performed to the highest order, namely cubic order in the
respective expansion variables $\Delta_F$ and $\Delta_{A_2}$, for
which the necessary inputs are available. We have taken the special
case $N_c=4$ and compared the results with values of these anomalous
dimensions in an SU(4) theory with $N_F=4$ and $N_{A_2}=4$ from a
lattice simulation in \cite{fx444}. We find agreement with these
measured values at the cubic order to which we have performed the
perturbative calculations, and we have given estimates of higher-order
corrections to our results. More generally, we have studied the dependence
of $\gamma_{\bar\psi\psi,IR}$ and $\gamma_{\bar\chi\chi,IR}$ as functions
of $N_F$ and $N_{A_2}$ in the SU(4) theory and have compared different
ways of calculating the lower boundary of the conformal window. 

% =======================================================================

\begin{acknowledgments}

  We thank Professors Yigal Shamir and Jong-Wan Lee for valuable
  discussions via email.  This research of R.S. was supported in part
  by the U.S. NSF Grant NSF-PHY-22-15093.

\end{acknowledgments}

% =======================================================================
% ======================================================================

\begin{appendix}

\section{Group Invariants}
\label{groupinvariants}

In this appendix we identify our notation for various group invariants.  
Let $T^a_R$ denote the generators of the Lie algebra of a group $G$ in the
representation $R$, where $a$ is a group index, and let $d_R$ denote the
dimension of $R$. 
The Casimir invariants $C_2(R)$ and $T_R$ are defined as follows:
$T^a_RT^a_R = C_2(R) I$, where here $I$ is
the $d_R \times d_R$ identity matrix, and
${\rm Tr}_R(T^a_R T^b_R) = T(R)\delta^{ab}$.  For a fermion $f$
transforming according to a representation $R$, we often use the equivalent
compact notation $T_f \equiv T(R)$ and $C_f \equiv C_2(R)$. We also use the
notation $C_A \equiv C_2(A) \equiv C_2(G)$.  Thus, e.g., 
for the $F$ and $A_2$ representations of SU($N_c$), $T(F)=1/2$,
$C_2(F)=(N_c^2-1)/(2N_c)$, $T(A_2)=(N_c-2)/2$, and
$C_2(A_2)=(N_c-2)(N_c+1)/N_c$. 

The coefficients $\kappa^{(f)}_j$ with $j \ge 3$ also involve higher-order
group invariants. In general, for a given representation $R$ of $G$,
\beqs
d^{abcd}_R &=&
\frac{1}{3!} {\rm Tr}_R\Big [ T^a(T^bT^cT^d + T^bT^dT^c + T^cT^bT^d \cr\cr
&+& T^cT^dT^b + T^dT^bT^c + T^dT^cT^b) \Big ] \ .
\label{d_abcd}
\eeqs
In \cite{dexm} we use the notation $d_R^{abcd} \equiv d_f^{abcd}$ and for
$R=Adj$, we write $d_R^{abcd}=d_A^{abcd}$. The $\kappa^{(f)}_3$
coefficients contain dependence upon products of these $d_R^{abcd}$ of
the form $d_R^{abcd}d_{R'}^{abcd} \equiv d_f^{abcd}d_{f'}^{abcd}$,
summed over the group indices $a, \ b, \ c, \ d$.  For further details
on these higher-order group invariants, see \cite{rsv} and references
therein.

\end{appendix}

% =======================================================================


\begin{thebibliography}{99}
\bibitem{fm}
Taking the fermions to be massless does not incur any loss of generality
because a fermion with a nonzero mass $m_0$ would be
integrated out of the low-energy effective field theory at Euclidean momentum 
scales $\mu < m_0$, and hence would be irrelevant to the properties of the
theory at the IRFP of interest here.
\end{thebibliography}
\end{document}